\documentclass{article}
\usepackage{epsf,latexsym}

\newcommand{\cqg}{Class. Quantum Grav.}
\newcommand{\jmp}{J. Math. Phys.}
\newcommand{\pr}{Phys. Rev. D}

\begin{document}

\title{\bf Constraints on Area Variables in Regge Calculus}

\maketitle
\author{Jarmo M\"akel\"a$^{a)}$ and Ruth M. Williams$^{b)}$  \\
                \\
$^{a)}$ Department of Physics, University of Jyv\"askyl\"a,  \\
P.O.Box 35, FIN-40351, Jyv\"askyl\"a, Finland.  \\
$^{b)}$ Girton College, Cambridge CB3 0JG, and   \\
DAMTP, CMS, Wilberforce Road, Cambridge CB3 0WA,  \\
United Kingdom, \\
California Institute of Technology,  \\
Pasadena, CA 91125, USA,  \\
and  \\
CIT-USC Center for Theoretical Physics,  \\
University of Southern California, Los Angeles,  \\
CA 90089-2535, USA.}


\section*{Abstract}
We describe a general method of obtaining the constraints between
area variables in one
approach to area Regge calculus, and illustrate it with a                      simple example.  The
simplicial complex is the simplest tessellation of the 4-sphere. 
The number of independent constraints on the variations of the 
triangle areas is shown to equal the difference between the numbers
of triangles and edges, and a general method of choosing these independent
constraints is described.  The constraints chosen by using our method
are shown to imply the Regge equations of motion in our example.

\section*{ }

Area Regge calculus \cite{rovelli} is a variant of conventional Regge
calculus in four dimensions, in which the triangle areas rather than
the edge lengths are used as dynamical variables \cite{barrett}.
Although for a single 4-simplex, the numbers of edges $N_1$ and of
triangles $N_2$ are equal, this is not true for a general simplicial
complex, where we usually have $N_2 > N_1$.  The counting of
degrees of freedom in a discrete theory is never completely
straightforward, and given the ambiguity in the number of true
variables, there are two attitudes one can take in area Regge
calculus.
\vskip15pt
Firstly, one can take the area variables as the fundamental ones, with
edges entering only as a tool for calculating volumes and deficit
angles.  (Recall \cite{barrett} that for a generic 4-simplex, it is
possible to solve uniquely for the edge-lengths in terms of the
areas.)  This approach has been investigated in detail in
\cite{barrett} and it has been shown \cite{williams} that although the
equations of motion constrain all the deficit angles to be zero, the
theory has non-trivial dynamical content.
\vskip15pt
Secondly, one can regard some of the areas as redundant variables and
aim to reduce their number to the number of edge lengths in the
simplicial complex.  This possibility has been considered in detail by
M\"akel\"a \cite{makela}.  In order to recover the conventional
approach to simplicial gravity where the edge lengths are real
physical quantities, it is necessary to impose the condition that a
given edge has the same length in whichever 4-simplex that length is
calculated.  (Strangely enough, this is by no means automatic - see 
\cite{barrett}.)  This leads to a large number of constraints: for
each edge, there is a constraint for each pair of 4-simplices sharing
that edge.  A total of $N_2 - N_1$ of these constraints will be
independent, but it is not easy to give a universal rule to say how to
choose them.  M\"akel\"a \cite{makela} has shown that if the
variations of the constraints are added in with Lagrange multipliers
to the variation of the Regge action expressed in terms of area
variables, then the usual Regge calculus equations of motion are
recovered. 
\vskip15pt
The purpose of this letter is to propose a general method of finding an
independent set of constraints and to illustrate it with a specific model.    
\vskip15pt
The key point of the method is the observation that when 2 4-simplices are
put together along a common tetrahedron, the number of edges is 14, whereas
the number of triangles is 16.  Thus the difference between the numbers of
triangles and edges is 2, from which it follows that we need 2 constraints
on the triangle areas to ensure that the edge-lengths in the tetrahedron
have consistent values, independent of the 4-simplex in which they are
calculated.  It appears that we can choose any 2 of the edges of the common
tetrahedron and that the corresponding constraints are always independent.
\vskip15pt
We now proceed further.  Denoting the 2 4-simplices already discussed above
by $\sigma_1$ and $\sigma_2$, we now consider a new 4-simplex $\sigma_3$,
which shares a common tetrahedron with $\sigma_2$.  This new 4-simplex
contains a number of new triangles, which is always greater than, or the
same as, the
number of new edges (the difference is usually, but not always, equal to
2).  We now pick a number, equal to this difference, of edges from the
tetrahedron shared between $\sigma_2$ and $\sigma_3$, and write down the
corresponding constraints.
\vskip15pt
This process is repeated.  We first count the difference between the number
of triangles and the number of edges appearing when a new 4-simplex is 
considered, and write the corresponding number of constraints, chosen as
described, until eventually the total number of constraints equals the
difference between the number of triangles and the number of edges for the
whole complex.   
\vskip15pt
We now describe a simple model to which we can apply this procedure.  The      simplicial
complex chosen is $\alpha_5$, the tessellation of $S^4$ by the surface
of a five-simplex, which is the complete graph on six points, labelled
0,1,...,5.   The numbers of simplices of each dimension are as
follows: $N_0=6, N_1=15, N_2=20, N_3=15, N_4=6$.  This means that we
expect the number of independent constraints to be $N_2-N_1=5$.
\vskip15pt
For simplicity, we start with all the edge lengths set equal to 1, and
the triangle areas to ${\sqrt{3}}/2$, and then allow small
variations.  We label the variations in squared-edge-length by                 $\delta s_{ij}$ and
in area by $\delta A_{ijk}$, where $i,j,k...$ label vertices.  Since
these variations are small, we keep only linear terms. 
\vskip15pt
To show how the constraints are derived, we give one example.  Solving
for the variation of the edge $s_{01}$ in the 4-simplex with
vertices 01234, we obtain 

\[
{(\sqrt{3}/2)} \delta s_{01}
=
2\delta A_{012} + 2\delta A_{013} + 2\delta A_{014} + 2\delta A_{234}
- \delta A_{023} - \delta A_{024} - \delta A_{034} 
\]
\begin{equation}
{- \delta A_{123} - \delta A_{124} - \delta A_{134}}.
\end{equation}

\noindent Similarly, in simplex 01235, we have 

\[
{(\sqrt{3}/2)} \delta s_{01}
=
2\delta A_{012} + 2\delta A_{013} + 2\delta A_{015} + 2\delta A_{235}
- \delta A_{023} - \delta A_{025} - \delta A_{035}
\]
\begin{equation}
{- \delta A_{123} - \delta A_{125} - \delta A_{135}}.
\end{equation}

\noindent Since we require $\delta s_{01}$ to be independent of the simplex in
which it is calculated, we obtain our first constraint by equating the
expressions in (1) and (2), giving

\[
2\delta A_{014} + 2\delta A_{234} - \delta A_{024} - \delta A_{034} - 
\delta A_{124} - \delta A_{134}
\]
\begin{equation}
{= 2\delta A_{015} + 2\delta A_{235} - \delta A_{025} - \delta A_{035} - 
\delta A_{125} - \delta A_{135}}.
\end{equation}
 
\noindent We denote this constraint by $[01,45]$.  In general, $[ij,kl]$ stands 
for the constraint that $\delta s_{ij}$ be the same in the two
4-simplices which have all vertices except $k$ and all except $l$.  
Clearly this makes sense only if neither $i$ nor $j$ is equal to $k$
or $l$.
\vskip15pt
Each edge is shared by four 4-simplices, so taking the expressions 
for $\delta s$ two at a time, we obtain 6 constraints for each edge. 
There are 15 edges, so the total number of constraints is 90.  However
this can be reduced to half immediately by noticing that, for each edge,
only 3 of the possible differences between pairs are independent.  (In 
general, if $n$ 4-simplices meet on an edge, then by equating the values
of the square of that edge-length in successive 4-simplices, we see that 
there will always be a total of $n-1$ independent constraints there.) 
This can be expressed symbolically by

\begin{equation}
{[ij,kl] = [ij,km] - [ij,lm]},
\end{equation}

\noindent where $m$ is the label of one of the remaining vertices.  Note that 
there is an ambiguity in the sign of each term, since the actual 
constraint is $[ij,kl]=0$.
\vskip15pt
By inspection of the constraints obtained, it can be seen that 

\begin{equation}
{[ij,kl] = [mn,kl]},
\end{equation}

\noindent where $(ijklmn)$ is one permutation of $(012345)$, and that

\begin{equation}
{[ij,kl] = [im,kl] - [jm,kl]},
\end{equation}

\noindent where again $m$ is the label of one of the remaining vertices.
\vskip15pt
The number of relations of the type given in (4) is 60, there  are
45 of type (5) and 60 of type (6), so clearly they are not all 
independent and we cannot subtract their total number from the 
number of constraints.
\vskip15pt
To understand the origin of the constraints of type (5), we need to 
consider the action of the symmetry group of the simplicial complex,
which is $S_6$, the permutation group on 6 objects.  For example,
consider the relation

\begin{equation}
{[01,45] = [23,45]}.
\end{equation}

\noindent Now the permutations in $S_6$ which exchange the edges $01$ and $23$
are $(0213)$, $(0312)$, $(02)(13)$ and $(03)(12)$.  Acting on 
triangles, they leave invariant the sets $(014,234)$, $(015,235)$,
$(012,013,023,123)$, $(024,034,124,134)$ and $(025,035,$ $125,135)$.  
These are precisely the sets of triangles which leave the
constraints $[01,45]$ and $[23,45]$ unchanged.
\vskip15pt
For the relations of type (6), the symmetries are more subtle; there
is an elaborate matching between the terms which enter the
constraints, and this is a feature of the particular simplicial 
complex used.
\vskip15pt
A convenient way of considering the constraints is as follows.  
Since there are 20 triangles, we regard the variations of areas (the
$\delta A_{ijk}$) as normalised basis vectors of a 20-dimensional 
vector space, $V_{20}$.  Each of the 90 constraints corresponds to a 
vector ${\bf C}_i, i=1,...,90$, in this space.  Clearly there cannot be
more than
20 linearly independent ${\bf C}_i$, and in fact we have shown by
straightforward but tedious calculation that only 5 of them are 
linearly independent. This is precisely the number $N_2 - N_1$ of
linearly independent constraints that we expect.
\vskip15pt
To see why this number has to be 5, consider a linear extension of
$V_{15}$, the 15-dimensional vector space with basis the normalised 
vectors corresponding to the variations of the edge lengths in the
simplicial complex.  We denote the vector corresponding to the
variation in edge $ij$ by $\bf v_{ij}$, and that corresponding to  
the variation in the area of triangle $ijk$ by $\bf V_{ijk}$.  We
define a linear homomorphism $\phi$ from $V_{15}$ into $V_{20}$ by

\begin{equation}
{\phi: \bf v_{ij} \rightarrow \bf V_{ijk} + \bf V_{ijl} + \bf V_{ijm}
+ \bf V_{ijn}},
\end{equation}

\noindent where $(ijklmn)$ is some permutation of $(012345)$.  This mapping is
$S_6$-linear:

\begin{equation}
{\phi(g {\bf v}) = g(\phi {\bf v}) \  for \  all \  g \in S_6},
\end{equation} 

\noindent and it is also injective, as can be seen by checking that its kernel
is trivial.  The vector space structure of $V_{15}$ is obviously
inherited by its image in $V_{20}$, so we have a 15-dimensional
subspace of $V_{20}$, which we denote by $Im(\phi)$.  This is an $S_6$-
invariant subspace of $V_{20}$.  It is then a 
simple matter to check that every vector $\bf C_i$ is orthogonal to
all the vectors spanning $Im(\phi)$.  For example, look back at
(2) and represent it by $\bf C_1$.  Then $\bf C_1$ is orthogonal to
$\bf V_{012} + \bf V_{013} + \bf V_{014} + \bf V_{015}$ and so on.
Thus the vectors $\bf C_i$ span a nontrivial subspace of $V_{20}$ orthogonal
to $Im(\phi)$, which must therefore have dimension at most 5.  It must also    be an $S_6$-invariant subspace of $V_{20}$ 
(note that the scalar product is invariant under $S_6$) and so must have
dimension precisely 5 (the 
representation theory of $S_6$ implies that the orthogonal complement
subspace is irreducible). 

\vskip15pt
We are now ready to apply our general method of choosing independent
constraints to our simplicial complex, $\alpha_5$.  Since it has only 6
vertices, it is convenient to denote the 4-simplex from which vertex $k$
is missing by $\sigma_k$.
\vskip15pt
Consider first the 4-simplices $\sigma_0$ and $\sigma_5$, which meet on
the tetrahedron with vertices (1234).  We pick the edges (12),(13), say,
from this tetrahedron, and write the corresponding constraints, which, in
the notation introduced above, are [12,05] and [13,05].  It is easily
checked that these are independent.
\vskip15pt
Next we consider the 4-simplex $\sigma_1$.  This meets $\sigma_0$ on the
tetrahedron (2345) and $\sigma_5$ on the tetrahedron (0234).  It brings in
one new edge (05) and 3 new triangles (025),(035),(045).  We choose 2 edges,
(23),(24) say, from the common tetrahedra, and write down 2 constraints for 
them, [23,01] and [24,01] say.  These are again independent of each other,
and also independent of the constraints already introduced.
\vskip15pt
So far we have constrained the variables in $\sigma_0$, $\sigma_1$ and 
$\sigma_5$, and we have 4 constraints out of the 5 that we need.  Let us
now consider $\sigma_2$, which meets $\sigma_1$ on the tetrahedron (0345).
We pick just one edge, (03) say, with the constraint [03,12].  It is
straightforward to check that this is independent of the others.  One  
possible set of constraints is then

\begin{equation}
{[12,05], [13,05], [23,01], [24,01], [03,12]}.
\end{equation}

\noindent Obviously the choice is not unique. 
\vskip15pt

Having obtained an independent set of constraints, we can then check that,
as in \cite{makela}, where all the constraints are added with Lagrange
multipliers to the action, we obtain the usual Regge equations of motion.
Suppose we take the set of constraints listed in (10), and add them to
the variation of the action, with Lagrange multipliers $\lambda_i$, giving
\cite{makela}

\[
\delta S = - {1\over {8\pi}} {1\over {3!}} \delta A_{ijk} \phi_{ijk}
\]
\begin{equation}
{+ \lambda_1 [12,05] + \lambda_2 [13,05] + \lambda_3 [23,01] +
\lambda_4 [24,01] + \lambda_5 [03,12]},
\end{equation}

\noindent where $\phi_{ijk}$ is the deficit angle at
the triangle with the corresponding labels.  Setting $\delta S = 0$
and equating coefficients of each $\delta A$, we obtain expressions for 
the deficit angles.  For example,

\begin{equation}
{\phi_{012} = 8 \pi (2\lambda_1 - \lambda_2)},
\end{equation}

\begin{equation}
{\phi_{013} = 8 \pi (- \lambda_1 + 2\lambda_2 + 2\lambda_5)},
\end{equation}

\begin{equation}
{\phi_{014} = 8 \pi (- \lambda_1 - \lambda_2 - \lambda_5)},
\end{equation}

\begin{equation}
{\phi_{015} = 8 \pi (- \lambda_5)}.
\end{equation}

\noindent Now the equation of motion obtained by varying with respect
to $s_{01}$ involves the sum over triangles of ${({\partial A_{ijk}}/ 
{\partial s_{01}})} \phi_{ijk}$.  Since the lattice we are using is totally
symmetric, the ${\partial A}/{\partial s}$ is a fixed number, independent of
both the triangle and of the edge, and the sum
we want just involves multiplying equations (15)-(18) by that number and 
adding them.  (Note that the deficit angles listed are for all the 
triangles which contain $s_{01}$.)  It is clear that we immediately
obtain the usual equation of motion

\begin{equation}
{{{\partial A_{ijk}} \over {\partial s_{01}}} \phi_{ijk} = 0}.
\end{equation}

\noindent Similarly for variations with repect to all the other $s_{ij}$.

\vskip 15pt

In this letter we have described a method of picking a set of 
independent constraints between variations of area variables in a 
general simplicial lattice.  Using this method we obtained a set of 
independent constrants in a simple example and showed that they imply
the ususal Regge equations of motion.  
This understanding of the general constraint structure of classical
area Regge calculus should pave the way for an investigation of the
corresponding quantum theory.

\section*{Acknowledgements}

The authors thank Jan Saxl for help with group theory.  One of them 
(RMW) is grateful for hospitality at the Theory Division at CERN, 
where this work was completed.  The work was supported in part by
the UK Particle Physics and Astronomy Research Council.

\end{document}